\documentstyle[aps,epsf,multicol]{revtex}
\begin{document}
\draft \widetext \title{Colored noise in the fractional Hall effect: duality
  relations and exact results}
\author{Claudio Chamon$^1$ and Denise E. Freed$^2$}
\address{$^1$ Department of Physics, Boston University, Boston, MA 02215}
\address{$^2$ Schlumberger - Doll Research, Old Quarry Road, Ridgefield, CT
06877}
\maketitle
\begin{abstract}
  We study noise in the problem of tunneling between fractional quantum Hall
  edge states within a four probe geometry. We explore the implications of
  the strong-weak coupling duality symmetry existent in this problem for
  relating the various density-density auto-correlations and
  cross-correlations between the four terminals. We identify correlations
  that transform as either ``odd'' or ``anti-symmetric'', or ``even'' or
  ``symmetric'' quantities under duality.  We show that the low frequency
  noise is colored, and that the deviations from white noise are exactly
  related to the differential conductance. We show explicitly that the
  relationship between the slope of the low frequency noise spectrum and the
  differential conductance follows from an identity that holds to {\it all}
  orders in perturbation theory, supporting the results implied by the
  duality symmetry. This generalizes the results of quantum supression of the
  finite frequency noise spectrum to Luttinger liquids and fractional
  statistics quasiparticles.
\end{abstract}

\pacs{PACS: 73.40.Hm, 71.10.Pm, 73.40.Gk}


\begin{multicols}{2}
\narrowtext

\section{Introduction}
\label{sec-intro}

\begin{figure}
\center
\noindent
\epsfxsize=1.8in
\epsfbox{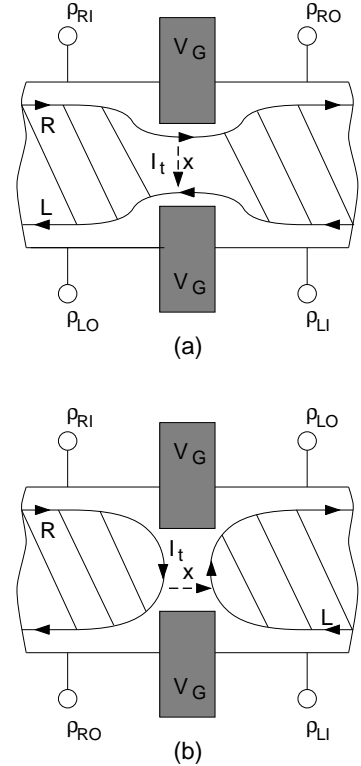}
\vspace{.5cm}
\caption{Two geometries for tunneling between edge states.  In (a)
quasiparticles can tunnel from one edge to the other. In (b) only
electrons can tunnel across. The subscripts for the densities $\rho$
in the four terminals determine the chirality ($R,L$) of the branch
and whether the branch is incoming to ($I$) or outgoing from ($O$) the
tunneling point.}
\label{fig1}
\end{figure}

Measurements of current fluctuations in a system can yield much
information about its excitation spectrum. This has been shown to be
the case for tunneling between edge states of fractional quantum Hall
(FQH) liquids. Two experimental groups, one in Saclay \cite{Glattli}
and another at the Weizmann Institute \cite{Reznikov1}, have recently
been able to measure the shot noise level in the tunneling current
between FQH edges. The results of the experiments are consistent with
the interpretation of tunneling of fractionally charged
quasiparticles. The geometries for such measurements are shown in Fig.
\ref{fig1}, and studies of various properties of the noise spectrum
have been carried out recently
\cite{K&F-noise,FLS1,FS2,LS3,CFW1,CFW2,Weiss}.

For a small tunneling current $I_t$ between the FQH liquid edges, the
shot noise level should approach the classical limit $2e^*I_t$, where
$e^*$ is either the Laughlin quasiparticle charge ($e^*=\nu e$) for
the geometry in Fig. \ref{fig1}a, or the electron charge ($e^*=e$) for
the geometry in Fig. \ref{fig1}b \cite{K&F-noise,FLS1,CFW1}. In this
classical limit, the tunneling events are uncorrelated and the noise
spectrum at low frequencies appear to be white or frequency
independent. As the tunneling current increases, two different effects
become manifest. First, the zero-frequency level deviates from the
classical level \cite{FLS1,FS2,Weiss}. Secondly, the low frequency
spectrum is no longer white, and develops a cusp at zero frequency
\cite{LS3,CFW1}. This colored noise structure offers the possibility
to investigate how the results known for non-interacting particles
with Fermi statistics are modified by the correlation effects in FQH
liquids, which contain excitations with fractional charge and
statistics.

It has been known for a while that statistics play a role in the
suppression of shot noise. In the case of transmission of electrons
through a quantum point-contact (QPC), with transmission coefficient
$T$, the zero frequency shot noise level is given by $2e I (1-T)$,
where $I=\frac{e^2}{h}T\;V$ is the current transmitted across the
point contact. The effects of the fermionic statistics is to reduce by
a factor $1-T$ the classical shot noise level $2e I$
\cite{Martin}. The implications of quantum statistics, however, are
not limited solely to the zero frequency noise level. The noise
spectrum is not white, so that the zero-frequency noise level alone
cannot describe the full frequency dependent noise spectrum.

For non-interacting electrons, the excess noise spectrum $S_{\rm
ex}(\omega)$, defined as the difference between the non-equilibrium
($V\ne 0$) and equilibrium ($V=0$) noise, is given by
\begin{equation}
S_{\rm ex}(\omega)={e^2\over \pi}\ T(1-T)\ (\omega_J-|\omega|)\
\theta(\omega_J-|\omega|)\ ,
\end{equation}
where $\omega_J=eV/\hbar$ is the ``Josephson'' frequency set by the
applied voltage. Notice that the excess noise decreases linearly with
frequency from the zero-frequency shot noise level $S_{\rm ex}(0)=2e I
(1-T)$ to zero at the Josephson frequency $\omega_J$, remaining zero
beyond this frequency scale.  Although it is hard to probe
experimentally the noise spectrum both at high-frequencies near
$\omega_J$ and at low-frequencies (due to $1/f$ noise), it is possible
to observe the spectrum at intermediate frequencies. Indeed,
measurements of the spectrum in this intermediate range have been
recently obtained by Reznikov {\it et. al.}\cite{Reznikov2}. These
measurements can be extrapolated to zero frequency, yielding the
experimental observation of the quantum suppression of shot noise in a
QPC by a factor $1-T$. Hopefully, further refinements of the
experimental technique would allow the measurement of the slope
${\Delta S_{\rm ex}\over \Delta \omega}=-{e^2\over \pi}\ T(1-T)$ of
the excess noise spectrum, thus probing quantum effects for finite
frequencies and showing that the excess noise spectrum is not white.

It is noteworthy that the slope of the noise spectrum near zero
frequency keeps a close relationship to the transport coefficient
$T$. This relationship is {\it not} the one supplied by the
fluctuation dissipation theorem (notice that the slope dependence on
$T$ is quadratic). One can also find the relationship between $T$ and
the slope of the full noise spectrum $S(\omega)=S_{\rm ex}(\omega)+
S_{\rm eq}(\omega)$ (excess plus equilibrium contribution to noise);
using $S_{\rm eq}(\omega)={e^2\over \pi}T\ |\omega|$, one finds that
\begin{equation}
  \label{eq:slope1}
  {\Delta S\over \Delta \omega}={e^2\over \pi}\ T^2 \ . 
\end{equation}

The relation connecting the slope of the noise spectrum at
low-frequencies to the transport coefficient $T$ through the point
contact, shown above for non-interacting electrons, can be
generalized. In this paper we will generalize such a relation to the
case of tunneling between chiral Luttinger liquids. These strongly
correlated states have excitations that carry fractional quantum
numbers, such as charge and statistics, which make them ideal
candidates for the study of the effects on quantum noise due to
generalized charge and statistics.

Chiral Luttinger liquids are realized on the edges of fractional
quantum Hall liquids. We will study the low-frequency slope of the
noise spectrum for a four probe geometry. This consists of looking at
density fluctuations in the left ($L$) and right ($R$) edges, both
incoming ($I$) to and outgoing ($O$) from the tunneling point. We will
show that correlations between densities in pairs of terminals are
related to the transmission and backscattering differential
conductances in ways dictated by the strong-weak duality symmetry
present in the problem.

The paper is organized as follows. In Section \ref{sec-results} we
summarize our results for the relationships between the slope of the
noise spectra and the differential conductances. In Section
\ref{sec-duality} we discuss the four terminal geometry for
measurement of auto and cross-correlations between pairs of terminals,
and we derive in detail the relationships for the correlations of
voltage/current fluctuations between pairs of terminals, which follow
from the dual descriptions of the problem in terms of electron or
quasiparticle tunneling (Fig. \ref{fig1}). By using current
conservation and another symmetry operation, which exchanges right $R$
and left $L$ edges and reverses the voltage, we are able to relate
correlations in one picture to those in the dual. We then show how
these relations tie, in particular, the slope of the spectra to
differential conductances in the problem. We derive these
relationships between noise spectra slope and differential
conductances directly from the boundary sine-Gordon model that
describes the tunneling problem in Sections \ref{sec:cross} (for
auto-correlations) and \ref{sec:current} (for the tunneling
current). We show that the relations hold to {\it all} orders in
pertubation theory, signaling the existence of an exact identity. We
conclude the paper in Section \ref{sec:conclusions} with a discussion
of our results, and a comparison to results on the low frequency
spectrum obtained from the thermal Bethe ansatz and form factors. In
the first reading of the paper, we suggest that readers peruse Section
\ref{sec:conclusions} prior to Sections \ref{sec:cross} and
\ref{sec:current}.

\section{Summary of results}
\label{sec-results}

\begin{figure}
    \begin{center}
      \noindent
      \epsfxsize=\linewidth \epsfbox{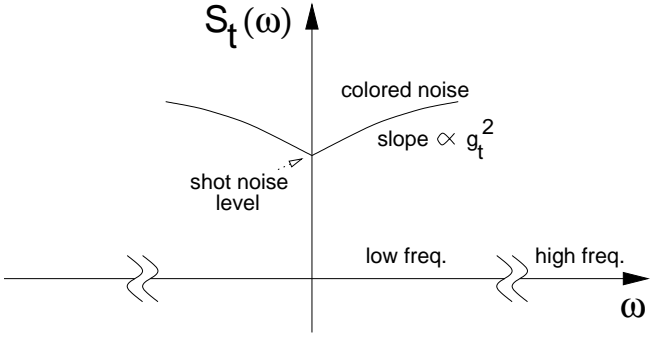} \hspace{1cm}
      \caption{Low frequency noise
        spectrum for the tunneling current $I_t$ between edges.}
      \label{fig:spectrum}
    \end{center}
\end{figure}

In this section we will give a brief summary of some of the main
results and ideas in this paper. We begin by describing the
low-frequency structure of the noise spectrum for the tunneling or
backscattering current $I_t$ between the edges of the FQH liquid, as
depicted in Fig.~\ref{fig:spectrum}.

The noise in the backscattering current is defined as
\[
S_t(\omega)=\int dt \; \cos(\omega t) 
\; \langle \{ I_t(t),I_t(0)\} \rangle \ \ .
\]
So far, much of the theoretical study of noise in the fractional
quantum Hall effect has focused on the zero-frequency shot noise level
or $S_t(\omega=0)$. For small tunneling currents the shot noise
approaches the classical level $S_t(\omega=0)=2 e^* I_t$. The solution
for the zero-frequency shot noise is known exactly, via the Bethe
ansatz, for any value of the applied voltage $V$ and tunneling
amplitude $\Gamma$. However, the finite $\omega$ results are rather
more complicated, and require knowledge of form factors in order to
calculate the current-current correlations defining the noise
spectrum~\cite{LS3}. In this paper we introduce an alternative
approach.

The spectrum at low frequencies is not white, {\it i.e.}, frequency
independent or flat. The frequency dependence or color of the noise
can be described in terms of the slope or derivative of the noise
spectrum with respect to the frequency:
\begin{equation}
A_t= \frac{1}{G_H}
\lim_{\omega\to 0} \frac{S_t(\omega)-S_t(0)}{\hbar |\omega|}\ ,
\end{equation}
where we have divided the slope by the Hall conductance $G_H=\nu
e^2/h$ in order to make $A_t$ a non-dimensional quantity.

In this paper we will show that the slope $A_t$ is directly related to the
differential backscattering conductance $G_t=d I_t/d V$ by
\begin{equation}
  \label{eq:slopeAt}
A_t = 2 \left({G_t \over G_H}\right)^2= 2\; g_t^2\ \ ,
\end{equation}
where we also define a dimensionless differential conductance
$g_t$. The relation in Eq. (\ref{eq:slopeAt}) is a generalization of
the result for non-interacting electrons (Luttinger parameter $g=1$)
to correlated chiral Luttinger liquids.  Similarly, we can relate the
noise in the transmission current, $S_T(\omega)$, to the differential
transmission conductance $G_T=d I_T/d V$ (with $g_T=G_T/G_H$):
\begin{equation}
A_T = 2\; g_T^2\ \ .  
\end{equation}

These results are valid in both the picture in Fig.~\ref{fig1}a, where
quasiparticles tunnel, or in the picture in Fig.~\ref{fig1}b, where
electrons tunnel. We indicate which picture we use by a $q$ or $e$
superscript in the quantity of interest, for example, $A_t^q,A_t^e$
and $g_t^q,g_t^e$.

This brings into focus another point we discuss in this paper: how
different quantities transform between the quasiparticle and electron
tunneling pictures. For example, it is widely known that
\begin{equation}
  \label{eq:dualg}
g_t^e=1-g_t^q \ ,  
\end{equation}
which can be interpreted as a duality relation. This comes about
because the two pictures describe the same physical system, but one is
the strong tunneling limit of the other and {\it
vice-versa}. Eq. (\ref{eq:dualg}) then simply follows from the
definitions of $I_t$ and $I_T$ in the two pictures of
Fig.~\ref{fig1}. These two pictures are actually dual in another,
stronger, sense: the model for the two pictures is self-dual,
{\it i.e.}, the tunneling mechanism for the two pictures is described
by the same Lagrangian, but with a different sets of parameters (up to
counterterms required for renormalization). The parameters are the
tunneling amplitude $\Gamma$ and the Luttinger parameter $g$. In
the quasiparticle tunneling picture they take the values $\Gamma_q$
and $g$, and in the electron tunneling picture they take the values
$\Gamma_q$ and $g^{-1}$, respectively. This allows us to write the
conductances in the two pictures using a single function $g_t$:
\begin{eqnarray}
g_t^q&=&g_t(g,\Gamma_q) \\
g_t^e&=&g_t(g^{-1},\Gamma_e)\ \ .
\end{eqnarray}

Because the backscattering conductances $g_t^q$ and $g_t^e$ in the two
pictures are related to each other by Eq.~(\ref{eq:dualg}) in such a
simple way, we can ask whether other physical quantities in the two
pictures have similar relationships.  The coefficients $A_t^q$ and
$A_t^e$ do not transform in a such a simple way.  Instead, we will
show that other quantities, such as the four terminal noise
correlations, do have simple transformation laws, although they do not
necessarily satisfy the same transformation rules as the conductance.
Under the duality transformation, cross-correlations between an
incoming and out-going branch transform in an ``odd'' or
``antisymmetric'' way like the backscattering conductance.  In
contrast, the auto-correlations transform in an ``even'' or
``symmetric'' way. Specifically, we show that the derivative with
respect to frequency of these noise correlations (as defined below)
satisfy
\begin{equation}
\label{eq:duala}
A_{\rm cross}^q=1-A_{\rm cross}^e
\qquad
A_{\rm auto}^q=A_{\rm auto}^e \ \ .
\end{equation}
We point out that $A_t$ and $A_T$ have both even and odd components,
which is why they do not transform in a simple way.

It is tempting to infer from Eqs.~(\ref{eq:dualg}) and
(\ref{eq:duala}) the ansatz $A_{\rm cross}=g_t$ or $A_{\rm
cross}=g_T$. (There are two combinations of cross correlations between
incoming and outgoing branches, depending on whether their chirality
is the same or opposite). These ansatz satisfy the correct
transformation under duality. Indeed, we show in this paper that this
ansatz is correct. This is supported by a perturbative calculation to
{\it all} orders. The calculation is done explicitly using the
self-dual boundary sine-Gordon theory that describes the tunneling
mechanism in both the quasiparticle and electron tunneling pictures.

The results for the slopes $A_{\rm cross}, A_t$ and $A_T$ can all be
expressed directly in terms of the differential conductance, bypassing
the non-linear series expansion in terms of the voltage $V$ and the
tunneling amplitude $\Gamma$. This is also the case for $A_{\rm
auto}$, which we find to be $A_{\rm auto}=1-2g_t
g_T=1-2g_t(1-g_t)$. In general, we can always write a quantity like
$A_{\rm cross}$ as a function $f(g,g_t)$ of both the Luttinger
parameter $g$ and the differential conductance $g_t$. However, our
results indicate that the differential conductance $g_t$ is the single
parameter controlling the behavior of the slope of the noise.

\section{Duality relations}
\label{sec-duality}

In this section, we will show how the symmetries of the system, which
include a voltage reversal symmetry and a duality symmetry, combined
with current conservation, lead to identities among the different
correlators.  We then make use of these identities and one further
ansatz about the dependence of the noise on the differential
conductance to solve for the noise in the Hall and tunneling currents
in terms of differential conductances.

We begin by describing two dual pictures of the system.  In the first
picture, the constriction is not pinched off, as in Figure
\ref{fig1}a, so the quasiparticles can tunnel from one edge to the
other.  The charge of the particle that tunnels is $e^*$. Its
tunneling amplitude is $\Gamma_{q}$, and the Luttinger parameter is
$g$. One may view $g$ as a parameter controlling the influence of a
tunneling event on subsequent ones.  In the second picture, the
constriction is completely pinched off, so there is no longer any
quantum Hall liquid in the central region.  Now only electrons can
tunnel from one edge to the other, so in this picture the charge of
the particle that tunnels is $e$. Its tunneling amplitude is
$\Gamma_e$, and the Luttinger parameter is $g^{-1}$.

There are two ways in which these pictures are dual to one another. We
will begin by describing the first one here, and save the second until
later in this section. The two pictures are dual to one another in the
sense that the strong tunneling limit of one should also describe the
weak tunneling limit of the other. For example, as the constriction in
Figure \ref{fig1}a is narrowed, it becomes easier for quasiparticles
to tunnel from one edge to another, so $\Gamma_q$ increases.  As the
constriction is narrowed further (so $\Gamma_q$ is increased some
more) at some point it pinches off completely, and we obtain the
second picture, Figure \ref{fig1}b, which is described by electron
tunneling with small $\Gamma_e$.  Thus the large $\Gamma_q$ limit of
the quasiparticle picture should be the same as the small $\Gamma_e$
limit of the electron picture, and {\it vice versa}.  In other words,
the two pictures should both describe the same physical system.

This means that the incoming and outgoing, right and left moving
densities in the two pictures of Fig.~\ref{fig1} are related by
\begin{eqnarray}
\rho_{RI}^q & = &  \rho_{RI}^e \label{rhodef} \\
\rho_{LI}^q & = &  \rho_{LI}^e \\
\rho_{RO}^q & = &  \rho_{LO}^e \\
\rho_{LO}^q & = &  \rho_{RO}^e ,
\label{rhodeffinal}
\end{eqnarray}
where the subscripts $I$, $O$ denote incoming and outgoing branches,
and $R$, $L$ denote right-moving and left-moving. The superscript $q$
or $e$ means the function for the density is given in the
quasiparticle or electron picture, respectively (see Figure
\ref{fig1}). As described above, the densities in the quasiparticle
picture are functions of $g$ and $\Gamma_q$ and the densities in the
electron picture are functions of $g^{-1}$ and $\Gamma_e$.

These equations can be used to relate the currents and correlations in
the quasiparticle picture to those in the electron picture. In order
to solve for these correlations, we need further constraints on them.
These constraints will take the form of a relation between the
correlator in one picture as a function of one set of parameters and
the same correlator in that picture as a function of another set of
parameters.  We will find that these relations will greatly restrict
the form of the correlators, but will not determine them uniquely.

To derive these relations, we must make use of additional properties
of the system. The first one is current conservation, which simply
states that, in a given picture,
\begin{equation}
\rho_{RO} + \rho_{LO} = \rho_{RI} + \rho_{LI}.
\end{equation}
(Strictly speaking, except right at the impurity, this equation is
purely classical. Once these densities appear in expectation values
for the noise, the correlators are modified by additional phases
$e^{i\omega x}$, where $x$ is the distance along the edge between the
terminal in question and the impurity). The second property is that
quantities that are quadratic in the densities, such as the noise, are
symmetric under inversion of applied voltage
\[
V \to -V \ .
\]
This implies that we can exchange the labels $R$ and $L$ without
changing the value of the density-density correlations. To see this,
just rotate the sample by 180 degrees, and invert the voltage: these
symmetry operations exchange the labels $R$ and $L$.

Next, it is useful to define some of the currents and noise
correlators in terms of the densities. The transmission current is
given by $I_T = \langle \rho_{RO} - \rho_{LI} \rangle $ and the
tunneling or backscattering current is given by $I_t = \langle
\rho_{RI} - \rho_{RO} \rangle $. It is also useful to define $I_H =
\langle \rho_{RI} - \rho_{LI} \rangle$, which is the Hall current in
the absence of tunneling or backscattering.

For the noise, we will define, for example,
\begin{equation}
  S_{RI, LO} (\omega)= \int dt\; cos(\omega t) \;
\langle \{\rho_{RI}(t) , \rho_{LO}(0)\}\rangle 
\end{equation}
The other noise correlators, for instance $S_{RO, LO}$ and $S_{RO,
RO}$, are defined likewise. (In general, we will denote by
$S_{\alpha,\beta}$ the correlator between the terminals
$\alpha,\beta$, where $\alpha$ and $\beta$ can take on the values
$RO,LO,RI,LI$).  Notice that we dropped the $x$ dependence of the
correlators. We do so because we are primarily interested in the
spectrum at low frequencies, in which case the $x$ dependence can be
neglected as long as $\omega \ll x^{-1}$. If necessary, to distinguish
which picture we are considering we will use the superscripts $e$ and
$q$.

We define the noise $S^{(0)}$ as the correlator between two right or
two left moving densities in the absence of the coupling $\Gamma$
between the $R$ and $L$ branches in a particular picture. It is given
by
\[
S^{(0)}(\omega)= \frac{\nu}{2\pi} |\omega|\ .
\]
We note that $S_{LI, LI} = S_{RI, RI}=S^{(0)}$ in the presence of any
tunneling because the incoming channels have yet to be affected by the
tunneling.  Similarly, $S_{LI,RI} = S_{RI, LI} = 0$ since the two
incoming channels are completely uncorrelated.

It follows from current conservation and the symmetry under voltage
inversion ($R\leftrightarrow L$) that, in a given picture,
\begin{eqnarray}
I_T & = & I_H - I_t \label{samepicdifnoise}\\
S_{RI, RO} & = &   S_{LI, LO} = S^{(0)}- S_{RI, LO}
   = S^{(0)}- S_{LI, RO} \\
S_{RO,RO} & = & S_{LO, LO} = S^{(0)}- S_{RO, LO} \ .
\label{samepicdifnoisefinal}
\end{eqnarray}
These equations relate one set of currents or noise in one picture to
another set of currents or noise in the same picture.  Next, we can
use the relations between the two pictures to write expressions for
the current or noise in one picture in terms of the same current or
noise found in the other picture.  Combining
Eqs.~(\ref{samepicdifnoise}-\ref{samepicdifnoisefinal}) and
(\ref{rhodef}-\ref{rhodeffinal}), we find
\begin{eqnarray}
&\ & I_t^q = I_H - I_t^e \\
&\ & S_{RI, RI}^e = S_{RI, RI}^q = S^0 = {\nu\over 2\pi} |\omega| \\
&\ & S^{(0)}-  S_{RI,LO}^q = S_{RI,LO}^e  \\
&\ & S_{RO,LO}^q = S_{RO,LO}^e.
\end{eqnarray}
Because these equations relate the noise in one picture to the noise
in the other picture, we cannot make further use of these equations
unless we either know what the noise is in one of the two pictures, or
we know another relation between the noise in the two different
pictures.  If we make an additional assumption, we can obtain this
second set of relations.

In particular, there is a second sense in which we mean the two
pictures are dual: we assume that the tunneling is described by a
self-dual theory, by which we mean that the description of the
tunneling mechanism is the same in both pictures.  Pictorially, what
this means is that we can reverse the shaded and unshaded regions in
Figure 1b so that it looks exactly like Figure 1a, rotated by 90
degrees.  This signifies that now in both pictures the tunneling
should be described in the same way, just with differing parameters --
$g^{-1}$, $e$, and $\Gamma_e$, or $g$, $e^*$ and $\Gamma_q$ --
depending on whether quasiparticles or electrons are tunneling.
(However, the filling fraction, $\nu$, of the shaded region remains
the same in both cases.)  By using the Luttinger liquid framework for
both the electron tunneling of Figure 1b and the quasiparticle
tunneling of Figure 1a, we are implicitly making this assumption.
(However, once questions of renormalization arise and counter-terms
must be added to one picture and not the other, it is no longer
guaranteed that the system really is described by a self-dual theory.)
Mathematically, in this framework we use the same Lagrangian, just
with different values of charge and tunneling amplitude.  Given this
second type of duality, we can replace $S_{\alpha, \beta}^e$ and
$S_{\alpha,\beta}^q$ by a single function,
$S_{\alpha,\beta}(g,\Gamma)$, so that $S_{\alpha,\beta}^e =
S_{\alpha,\beta}(g^{-1}, \Gamma_e)$ and $S_{\alpha,\beta}^q =
S_{\alpha,\beta}(g, \Gamma_q)$, (and similarly for the currents). The
identities between the noise and currents then become
\begin{eqnarray}
I_t(g, \Gamma_q) & = & I_H - I_t(g^{-1}, \Gamma_e) \label{eq:I-dual}\\
S^{(0)} - S_{RI, LO}(g, \Gamma_q) &=& S_{RI, LO}(g^{-1}, \Gamma_e) \\
S_{RO,LO}(g, \Gamma_q) & = & S_{RO,LO}(g^{-1},
\Gamma_e).\label{eq:Sc-dual}
\end{eqnarray}
Now each of these equations is a duality relation relating a function
at one set of parameters to the same function at another set of
parameters.  This kind of relation greatly restricts the possible form
of the current and noise.

To find the form of the noise that is suggested by these duality
relations, we begin by noting that Eq.~(\ref{eq:I-dual}) for $I_t$ and
Eq.~(\ref{eq:Sc-dual}) for $S_{RI,LO}$ have a very similar form, which
indicates there may be a simple relation between the function that
satisfies the duality relation for the current and the one that
satisfies the duality relation for the cross-correlation $S_{RI,LO}$.
One must take care, though, in trying to equate $I_t$ and $S_{RI, LO}$
because they have different dimensions and one is a function of
$\omega$ and the other is not.  Instead, we will look at the slope
$A_{RI, LO}$ of $S_{RI,LO}$ near $\omega = 0$, which we will normalize
as follows:
\begin{equation}
A_{RI, LO} = \lim_{\omega\rightarrow  0} {S^{\rm sing}_{RI, LO} \over
S^{(0)}}.
\end{equation}
In this equation, $S^{\rm sing}_{RI, LO}$ is the part of $S_{RI, LO}$
that is singular as $\omega \rightarrow 0$.  We will also define the
dimensionless differential conductance as
\begin{equation}
g_t = {1 \over G_H} {dI_t \over dV},
\end{equation}
where $G_H = \nu e^2/h$.  
Then the duality relations for $A_{RI, LO}$ and $g_t$
become
\begin{eqnarray}
g_t(g, \Gamma_q) & = & 1 - g_t(g^{-1}, \Gamma_e) \\
A_{RI,LO}(g, \Gamma_q) & = & 1 - A_{RI,LO}(g^{-1}, \Gamma_e).
\end{eqnarray}
Thus the dimensionless conductance and the normalized slope of the
noise satisfy exactly the same duality relation.

To look for a simple relation between the conductance and $A_{RI,
LO}$, we consider the case when the Luttinger parameter $g = 1/2$.  In
this case, there is an exact solution for the noise.  Guided by the
fact that the conductance and the slope of the noise satisfy the same
duality relation, we find we can rewrite the expression for $A_{RI,
LO}$, calculated in reference \cite{CFW2}, in terms of the
conductance.  It is given by
\begin{equation}
A_{RI,LO} = g_t.
\label{ARILOgt}
\end{equation}
Thus we see that in this case there is, indeed, a simple relation
between $A_{RI, LO}$ and the differential conductance, which begs the
question of whether this is true for all $g$.  In Section
\ref{sec:cross}, we will explicitly calculate the slope of the noise
to all orders in perturbation theory.  This perturbative expansion is
valid when $(V/T_B)^{(2g-2)}$ is small, where $T_B \propto
\Gamma^{-1/(g-1)}$.  However, if we want to know the value for
$(V/T_B)^{(2g-2)}$ large, we can use the duality relation and
calculate in the dual picture.  In this way, we obtain the value of
$A_{RI, LO}$ over the whole region of parameter space.  We find that
Equation (\ref{ARILOgt}) does hold for all $g$.

Here, instead, we will show how relation (\ref{ARILOgt}) follows from
one additional ansatz.  Coloumb gas expansions for the conductance and
noise are both power series in $(V/T_B)^{(2g-2)}$ beginning with the
term $(V/T_B)^{(2g-2)}$.  As a consequence, we can always express
$A_{RI, LO}$ as a power series in $g_t$, given by
\begin{equation}
A_{RI, LO} = \sum_{n=1}^\infty a_n(g) g_t^n,
\end{equation}
where $a_n(g)$ depends on the Luttinger parameter $g$.  Because both
$A_{RI, LO}$ and $g_t$ satisfy the duality relation, this puts some
restrictions on the $a_n$, but does not determine them uniquely.
However, if we make the assumption that $a_n$ does not depend on $g$,
then we can use the solution at $g = 1/2$ to fix the $a_n$, with the
result that $A_{RI, LO} = g_t$ for all $g$.  (This assumption is
equivalent to the one made in \cite{Weiss} that enabled Weiss to use
the duality relation to solve for the conductance.  Since $A_{RI, LO}$
satisfies the same duality relation as the conductance and equals the
conductance for $g = 1/2$, his ansatz and ensuing calculation also
uniquely determine $A_{RI, LO}$).

We can use the same line of reasoning for the slope of $S_{RO, LO}$,
which is defined as
\begin{equation}
A_{RO,LO} = \lim_{\omega \rightarrow 0} {S_{RO,LO}^{\rm sing} \over S^{(0)}}.
\end{equation}
For $g = 1/2$, this can be written in terms of the 
differential conductance as follows:
\begin{equation}
A_{RO, LO} = 2(g_t - g_t^2).
\label{AROLOgt}
\end{equation}
Given the duality relation for $g_t$, this expression satisfies the
duality relation for $A_{RO, LO}$.  As before, $A_{RO, LO}$ is a power
series in $(V/T_B)^{(2g-2)}$ starting with the first order term, so it
is given by
\begin{equation}
A_{RO, LO} = \sum_{n=1}^{\infty} b_n(g) g_t^n.
\end{equation}
If again we assume the $b_n$ are independent of $g$, then the value of
$A_{RO,LO}$ at $g = 1/2$ determines $b_n$ for all $g$, and the slope
of the noise has the form given in Eq.~(\ref{AROLOgt}).
Combining this result with Eq.~(\ref{samepicdifnoisefinal}), which
relates correlations between pairs of outgoing branches, we can write
$A_{RO,RO}=1-A_{RO,LO}$, or
\begin{equation}
A_{\rm auto}=A_{RO,RO}=1-2 g_t g_T \ .
\end{equation}
This is our conjecture for the auto correlations in the outgoing
branches.

Finally, we conclude this section by stating the conjecture for the
slope of the noise in the tunneling current, $A_t$, and in the
transmission current, $A_T$.  The slope of the tunneling noise is
defined as
\begin{equation}
A_t = \lim_{\omega\rightarrow  0} {S_{t} \over S^{(0)}},
\end{equation}
where $S^{\rm sing}_{t}$ is the singular part of the tunneling
noise, and $A_T$ is defined similarly.  Using the definition of the
transmission and tunneling currents, we find
\begin{equation}
A_t = -A_{RO,LO} + 2 A_{RI, LO}
\end{equation}
and
\begin{equation}
A_T = 2 - A_{RO,LO} - 2A_{RI, LO}
\end{equation}

The noise in the transmission current in one picture equals the noise
in the tunneling current in the other picture.  However, in a given
picture, there is no simple duality relation for $A_t$ or $A_T$.
Instead, our solutions for $A_{RO,LO}$ and $A_{RI, LO}$ imply that
\begin{eqnarray}
A_t & = & 2 g_t^2, \\
A_T & = & 2 g_T^2.
\label{AtgtATgT}
\end{eqnarray}
In Section \ref{sec:current}, we use a multipole expansion to
calculate the tunneling noise to all orders in perturbation theory,
and find that Equation (\ref{AtgtATgT}) does, indeed, hold.  Thus, we
expect to find a simple relation between the slope of the noise at low
frequency and the square of the differential conductance.  This has
the same form as Shiba's relation for the dissipative two-level
system \cite{Shiba,Sassetti&Weiss}.

\end{multicols}
\widetext

\section{Cross-correlations}
\label{sec:cross}
In the previous section we related the slopes of different noise
correlations to the differential conductances by exploring the duality
symmetry in the problem. We used an ansatz that the coefficients for
the series expansions of these slopes (as a power series in the
conductances) did not depend on the Luttinger parameter. In this
section we formally justify this ansatz by explicitly calculating,
starting from the boundary sine-Gordon Lagrangian, the noise
cross-correlations between an incoming branch and an outgoing
branch. In other words, here we show that the independence of the
expansion coefficients on the Luttinger parameter $g$ is a specific
property of the boundary sine-Gordon Lagrangian that describes the
tunneling problem.

The Lagrangian that describes the tunneling between chiral Luttinger
liquids through a QPC is
\begin{equation}
{\cal L}={\cal L}_R+{\cal L}_L+\Gamma\ \delta(x)\ e^{i\omega_J t}
\ \ e^{i\sqrt{g}(\phi_R(t,0)+\phi_L(t,0))}\ ,
\end{equation}
where ${\cal L}_{R,L}={1\over 4\pi}\partial_x\phi_{R,L}
(\mp\partial_t-\partial_x)\phi_{R,L}$ is the Lagrangian for the free
chiral bosons. We will calculate the zero-frequency singularity in the
noise spectrum to all orders in a perturbative expansion in $\Gamma$.

We will show that the slope of the spectrum for the cross-correlations
is linearly related to the differential transmission and
backscattering conductances. We proceed in the following way: we will
expand the density-density correlations for the cross noise to all
orders in the tunneling amplitude $\Gamma$, and compare them to the
expansion, also to all orders, of the differential conductances.

We write the correlations between density operators as follows:
\begin{equation}
\langle \rho_{a}(t,x_1) \rho_{b}(0,x_2) \rangle\ ,
\end{equation}
where $a,b$ take the values $+1$ for $R$ moving branches and $-1$
for $L$ moving ones. Such compressed notation makes it simpler to
identify incoming and outgoing branches in a unified way for both
left and right movers: $\rho_a(t,x_1)$, for example, is the density
in an incoming or outgoing branch if $a x_1<0$ or $a x_1>0$,
respectively.

\begin{figure}
\hspace{.625 in}
\center
\noindent
\epsfxsize=4in
\epsfbox{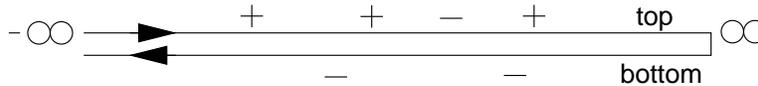}
\vspace{.5cm}
\caption{Keldysh contour for the non-equilibrium Coulomb gas
expansion. Time evolves forward in the top part of the contour, and
backwards in the bottom. Charges from the Coulomb gas expansion are
inserted in both the top and bottom pieces.}
\label{figcontour}
\end{figure}

The densities are related to the fields $\phi_{R,L}$ through
$\rho_{R,L}=\frac{\sqrt{\nu}}{2\pi}\partial_x \phi_{R,L}$, so that we
can write
\begin{equation}
\langle
\rho_{a}(t,x_1) \rho_{b}(t',x_2) \rangle =
\frac{\nu}{(2\pi)^2}\ \partial_{x_1}\partial_{x_2}\langle
\phi_{a}(t,x_1)\phi_{b}(t',x_2) \rangle\ ,
\end{equation}
where it is convenient to use
\begin{equation}
\langle \phi_{a}(t,x_1)\phi_{b}(t',x_2)
\rangle=
\frac{d\ }{d\lambda_1}\frac{d\ }{d\lambda_2}
\langle
e^{i\lambda_1\phi_a(t,x_1)}\ e^{-i\lambda_2\phi_b(t',x_2)}
\rangle{\big |}_{\lambda_1,\lambda_2=0}\ .\label{phiphi}
\end{equation}
The last correlation function is easy to calculate perturbatively
using
\begin{equation}
\langle T_c(e^{i\lambda_1\phi_a(t,x_1)}\ e^{-
i\lambda_2\phi_b(t',x_2)}) \rangle=
\langle 0|\ T_c(S(-\infty,-\infty)\ e^{i\lambda_1\phi_a(t,x_1)}\
e^{-i\lambda_2\phi_b(t',x_2)}) \ |0\rangle\ ,
\end{equation}
where $|0\rangle$ is the unperturbed ground state, and $T_c$ is the
ordering along the Keldysh contour (see Fig.~\ref{figcontour} and
Refs. \cite{CFW1,CFW2}). The scattering operator $S(-\infty,-\infty)$
takes the initial state, evolves it from $t=-\infty$ to $t=\infty$ and
back to $t=-\infty$. The use of the Keldysh contour is necessary in
the treatment of non- equilibrium problems, such as the one we have in
hand. A more detailed description of the method in the context treated
here can be found in Ref. \cite{CFW1}.

The perturbative treatment corresponds to a Coulomb gas expansion
\cite{CFW1}. The nonzero contribution to the correlation above comes
from the neutral terms in the expansion, thus only even orders in
$\Gamma$ contribute. To $(2n)$-th order, we have an insertion of $n$
positive charges and $n$ negative charges. We will label the times at
which they are inserted in the expansion $t_i$, $i=1,...,2n$, with
$i=1,\dots,n$ corresponding to the $+$ charges ($q_i=+1$), and
$i=n+1,\dots,2n$ for the $-$ charges ($q_i=-1$).
\begin{eqnarray}
\langle &T_c&(e^{i\lambda_1\phi_a(t,x_1)}\ e^{-
i\lambda_2\phi_b(t',x_2)}) \rangle=\\
&\ &\sum_{n=0}^\infty\ (i\Gamma)^n(i\Gamma^*)^n
\oint_c \prod_{i=1}^{2n} dt_i \ e^{i\sum_{i=1}^n \omega_0(t_i-t_{n+i})}
\langle 0|\ T_c\left(\ e^{i\lambda_1\phi_a(t,x_1)}\
e^{-i\lambda_2\phi_b(t',x_2)}
\ \prod_{i=1}^{2n} e^{i q_i \sqrt{g}\phi(t_i,0)}\right)
|0\rangle\ ,\nonumber
\end{eqnarray}
where $\phi$ without subscript stands for the sum $\phi_R+\phi_L$.
The expression above is simplified using
\begin{equation}
\langle 0|T_c (\prod_j e^{i q_j\;\sqrt{g}\phi(t_j,x_j)})|0 \rangle=
e^{-{g\over2}\sum_{i\ne j}q_iq_j
\langle
0|T_c(\phi(t_i,x_i)\phi(t_j,x_j))|0\rangle}\ .
\end{equation}
Substituting it into Eq. (\ref{phiphi}) we obtain
\begin{eqnarray}
\langle T_c(\phi_{a}(t,x_1)\phi_{b}(t',x_2))
\rangle&=&\sum_{n=0}^\infty\ (-1)^n\ |\Gamma|^{2n}
\oint_c \prod_{i=1}^{2n} dt_i 
\ e^{i\sum_{i=1}^n \omega_0(t_i-t_{n+i})}
e^{-{g\over2}\sum_{i\ne j}q_iq_j
\langle
0|T_c(\phi(t_i)\phi(t_j,x_j))|0\rangle}
\nonumber\\
&\ &\ \times 
\Bigg\{\left[\sum_{i=1}^{2n} \ q_i\;\sqrt{g}\; 
\langle 0|T_c(\phi(t_i,0)\phi_a(t,x_1))|0\rangle\right]
\times\left[\sum_{j=1}^{2n} \ q_j\;\sqrt{g}\; \langle
0|T_c(\phi(t_j,0)\phi_b(t',x_2))|0\rangle
\right]\nonumber\\
&\ &\ \ \ \ \ \ +
\langle 0|T_c(\phi_a(t,x_1)\phi_b(t',x_2))|0\rangle\Bigg\}
\ \label{correl}
\end{eqnarray}
The last term in the expression above is simply proportional to $\langle
0|T_c(\phi_a(t,x_1)\phi_b(t',x_2))|0\rangle$. The proportionality
constant is equal to $Z=\langle 0|S(-\infty,-\infty)|0\rangle\equiv
1$. This is the zero-order contribution.

In order to carry out the calculations, we introduce notation that
keeps track of the position of the inserted charges along the contour,
{\it i.e.}, whether they are in the forward (or top) branch, or in the
return (or bottom) branch (see Fig.~\ref{figcontour} and
Refs. \cite{CFW1,CFW2}). The position of the charges is important for
the computation of the contour-ordered correlation function
$\langle0|T_c(\phi_{R,L}(t_1,x_1)\phi_{R,L}(t_2,x_2))|0\rangle$
\begin{equation}
\ =\ 
\cases
{
-\ln \{\delta +i\ {\rm sign}(t_1-t_2)[(t_1-t_2)\mp(x_1-x_2)]\},&\mbox{
both $t_1$ and $t_2$ in the top branch}\cr
-\ln \{\delta -i\ {\rm
sign}(t_1-t_2)[(t_1-t_2)\mp(x_1-x_2)]\},&\mbox{ both $t_1$  and $t_2$
in the bottom branch}\cr
-\ln \{\delta -i[(t_1-t_2)\mp(x_1-x_2)]\},&\mbox{ $t_1$  in the top
and $t_2$ in the bottom branch}\cr
-\ln \{\delta +i[(t_1-t_2)\mp(x_1-x_2)]\},&\mbox{ $t_1$  in the
bottom and $t_2$ in the top branch.}\cr
}
\end{equation}
The compact notation consists of giving indices to the times which contain
the information about which branch of the Keldysh contour they are on,
so that $t^\mu$ is on the top branch for $\mu=+1$, and on the bottom for
$\mu=-1$. In this way, we can compress the correlations to a
compact form:
\begin{eqnarray}
G_{{\mu_1}{\mu_2}}^{ab}(t_1,x_1;t_2,x_2)&=&G_{{\mu_1}{\mu_2}}^{ab}(t_1-t_2,x
_1-
x_2)= \langle
0|T_c(\phi_a(t^{\mu_1},x_1)\phi_b(t^{\mu_2}_2,x_2))|0\rangle\nonumber\\
&=&- \delta_{a,b}\ \ln (\delta +i\ K_{{\mu_1}{\mu_2}}(t_1-t_2)[(t_1-t_2)-
a(x_1-x_2)])\ ,\label{G}
\end{eqnarray}
where
$K_{\pm\pm}(t)=\pm{\rm sign}(t)$ and $K_{\pm\mp}(t)=\mp 1$.
Again, we have used $a,b=\pm 1$ for $R$ and $L$ fields,
respectively. The correlation in Eq. (\ref{correl}) can be written,
using this compressed notation, as
\begin{eqnarray}
\langle T_c(\phi_{a}(t,x_1)\phi_{b}(t',x_2))
\rangle
&=&G_{++}^{ab}(t-t',x_1-x_2)\ +\nonumber\\
&\ &\!\!\!\!\!\!\!\!\!\!\!\!\!\!\!\!\!\!\!\!\!
\!\!\!\!\!\!\!\!\!\!\!\!\!\!\!\!\!\!\!\!\!
\!\!\!\!\!\!\!
g\;\sum_{n=1}^\infty\ (-1)^n\ |\Gamma|^{2n}
\sum_{\{\mu_i\}}\int_{-\infty}^{\infty} \prod_{i=1}^{2n}\ \mu_idt_i 
\ \ \ e^{i\sum_{i=1}^n \omega_0(t_i-t_{n+i})}
P_{{\mu_1},\dots,{\mu_{2n}}}(t_1,\dots,t_{2n})\nonumber\\
&\ &\ \ \ \times
\left[\sum_{i=1}^{2n} \ q_i\ 
G_{+{\mu_i}}^{aa}(t-t_i,x_1)\right]
\times
\left[\sum_{i=j}^{2n} \ q_j\ 
G_{+{\mu_j}}^{bb}(t'-t_j,x_2)
\right],
\label{corG}
\end{eqnarray}
where $P_{{\mu_1},\dots,{\mu_{2n}}}(t_1,\dots,t_{2n})
=e^{-{g\over2}\sum_{i\ne j}q_iq_j
[G_{{\mu_i}{\mu_j}}^{++}(t_i-t_j,0)
+G_{{\mu_i}{\mu_j}}^{--}(t_i-t_j,0)]}$.
The factors $\mu_i$ simply keep track of the sign coming from the
integration of the times $t_i$ along the contour. Notice that the
times $t$ and $t'$ are taken to be on the top branch.

Now, let
\begin{eqnarray}
F_{ab}(\omega;x_1,x_2)&=&\int_{-\infty}^\infty dt\
e^{i\omega t}\ \langle T_c(\rho_{a}(t,x_1)\rho_{b}(0,x_2))
\rangle\nonumber\\
&=&\frac{\nu}{(2\pi)^2}\ \partial_{x_1}\partial_{x_2}
\int_{-\infty}^\infty dt\
e^{i\omega t}\ \langle T_c(\phi_{a}(t,x_1)\phi_{b}(0,x_2))
\rangle\ ,
\end{eqnarray}
which can be easily shown, using Eq. (\ref{corG}), to yield
\begin{eqnarray}
F_{ab}(\omega;x_1,x_2)&=&
-\frac{\nu}{(2\pi)^2}\  \partial_{x_1}
{\tilde g}_{++}^{ab}(\omega,x_1-x_2)\nonumber\\
&+&
\frac{\nu g}{(2\pi)^2}
\sum_{n=1}^\infty\ (-1)^n\ |\Gamma|^{2n}
\sum_{\{\mu_i\}}\int_{-\infty}^{\infty} \prod_{i=1}^{2n}\ \mu_idt_i 
\ \ \ e^{i\sum_{i=1}^n \omega_0(t_i-t_{n+i})}
P_{{\mu_1},\dots,{\mu_{2n}}}(t_1,\dots,t_{2n})\nonumber\\
&\ &\ \ \ \times
\left[\sum_{i=1}^{2n} \ q_i\ 
{\tilde g}_{+{\mu_i}}^{aa}(\omega,x_1)\ e^{i\omega t_i}\right]
\times
\left[\sum_{j=1}^{2n} \ q_j\ 
g_{+{\mu_j}}^{bb}(t'-t_j,x_2)
\right]\ .
\label{F}
\end{eqnarray}
In this equation, the function $g$ is given by
$g_{{\mu_1}{\mu_2}}^{ab}(t,x)=\partial_{x}G_{{\mu_1}{\mu_2}}^{ab}(\omega,x)$
and ${\tilde g}$ is the Fourier transform of $g$; they can be obtained
from Eq.  (\ref{G}):
\begin{equation}
{\tilde g}_{{\mu_1}{\mu_2}}^{ab}(\omega,x)=\delta_{a,b}\times \ 
\cases
{
\pi ia\ e^{i\omega ax}\ \left({\rm sign}(\omega)+{\rm sign}(ax)\right)
&\mbox{, ${\mu_1}=+1,{\mu_2}=+1$}\cr
\pi ia\ e^{i\omega ax}\ \left({\rm sign}(\omega)-{\rm sign}(ax)\right)
&\mbox{, ${\mu_1}=-1,{\mu_2}=-1$}\cr 
-2\pi ia\ e^{i\omega ax}\ \theta(-\omega)
&\mbox{, ${\mu_1}=+1,{\mu_2}=-1$}\cr 
2\pi ia\ e^{i\omega ax}\ \theta(\omega)
&\mbox{, ${\mu_1}=-1,{\mu_2}=+1$}\cr
}
\end{equation}

We will now take $ax_1<0$ (incoming state) and $bx_2>0$ (outgoing
state), so as to discuss the cross-correlations. In this case,
${\tilde g}_{+\mu}^{aa}(\omega,x_1)=-2\pi ia e^{i\omega
ax_1}\theta(-\omega)$, for both $\mu=\pm1$. For small $|\omega|$ we
have
\begin{equation}
\sum_{i=1}^{2n} \ q_i\ 
{\tilde g}_{+{\mu_i}}^{aa}(\omega,x_1)\ e^{i\omega t_i}=
-2\pi a \ \theta(-\omega) \ \omega \sum_{i=1}^{n} (t_i-t_{n+i})
+{\cal O}(\omega^2)\ .
\end{equation}
One can thus write
\begin{eqnarray}
F_{ab}(\omega;x_1,x_2)&=&
-\delta_{a,b}\frac{\nu}{(2\pi)}\ \omega\ \theta({-\omega})\nonumber\\
&-&
a \ \theta(-\omega)\  \omega\  \frac{\nu g}{(2\pi)}
\sum_{n=1}^\infty\ (-1)^n\ |\Gamma|^{2n}
\sum_{\{\mu_i\}}\int_{-\infty}^{\infty} \prod_{i=1}^{2n}\ \mu_idt_i 
\ \ \ \sum_{i=1}^{n} (t_i-t_{n+i})\ 
e^{i\sum_{i=1}^n \omega_0(t_i-t_{n+i})}
\nonumber\\
&\ &\ \ \ \times\ P_{{\mu_1},\dots,{\mu_{2n}}}(t_1,\dots,t_{2n})
\left[\sum_{j=1}^{2n} \ q_j\ 
g_{+{\mu_j}}^{bb}(t'-t_j,x_2)
\right]
+{\cal O}(\omega^2)\ .
\label{Fomegato0}
\end{eqnarray}
We can similarly (and more easily) expand the tunneling current ({\it
i.e.}, the difference in the densities in a branch before and after
the impurity) to all orders in $\Gamma$.
\begin{eqnarray}
I_t=b\left(\langle\rho_{b}(t',x_2)\rangle 
- \langle\rho_{b}(t',-x_2) \rangle\right) &=& 
-ib\frac{\sqrt{\nu g}}{2\pi}
\sum_{n=1}^\infty\ (-1)^n\ |\Gamma|^{2n}
\sum_{\{\mu_i\}}\int_{-\infty}^{\infty} \prod_{i=1}^{2n}\ \mu_idt_i 
\ \ \  
e^{i\sum_{i=1}^n \omega_0(t_i-t_{n+i})}
\nonumber\\
&\ &\ \ \ \times\ P_{{\mu_1},\dots,{\mu_{2n}}}(t_1,\dots,t_{2n})
\left[\sum_{j=1}^{2n} \ q_j\ 
g_{+{\mu_j}}^{bb}(t'-t_j,x_2)
\right]\ .
\end{eqnarray}
The current is defined as positive flowing from the right to the left
edge, hence the factor $b$ in the expression above. By direct
comparison with Eq. (\ref{Fomegato0}), one can then write
\begin{equation}
F_{ab}(\omega;x_1,x_2)=
\left(-\delta_{a,b}\frac{\nu}{(2\pi)}\ 
+\sqrt{\nu g}\ ab\ \frac{dI_t}{d\omega_0}\right)
\omega\ \theta(-\omega)
+{\cal O}(\omega^2)
\end{equation}

The noise spectrum is obtained from
$F_{ab}(\omega,x_1,x_2)$ as follows:
\begin{eqnarray}
S_{ab}(\omega;x_1,x_2)&=&S_{ba}(-
\omega;x_2,x_1)=\int_{-\infty}^\infty dt\ e^{i\omega t} \langle
\{\rho_a(t,x_1),\rho_b(0,x_2)\}\rangle\nonumber\\
&=&F_{ab}(\omega;x_1,x_2)+F^*_{ab}(-\omega;x_1,x_2)\ ,\label{S}
\end{eqnarray}
so that we can finally write
\begin{eqnarray}
S_{ab}(\omega;x_1,x_2)&=&\left(\delta_{a,b}\frac{\nu}{(2\pi)}\ 
-\sqrt{\nu g}\ ab\ \frac{dI_t}{d\omega_0}\right) |\omega|
+{\cal O}(\omega^2)\nonumber \\
&=&\frac{\nu}{(2\pi)}|\omega| \left(\delta_{a,b}-ab\ {G_t\over G_H}\right)
+{\cal O}(\omega^2)\ ,
\end{eqnarray}
where $G_t={dI_t\over dV}$, {\it i. e.}, the differential conductance, and
$G_H=\nu e^2/h$ is the quantized Hall conductance.  We used above that
$\omega_0=e^* V/\hbar$, and that $e^*=\sqrt{\nu g}$ (in units of $e=1$).
Reinserting back $\hbar$ and $e$ (which were both set to 1), we can write the
result in a more physical way:
\begin{equation}
S_{ab}(\omega;x_1,x_2)=\hbar |\omega|\ \left(\delta_{a,b}\ G_H-ab\ G_t\right)
\ ,
\end{equation}
the final result for cross-correlations (valid when
$(ax_1)\times(bx_2)<0$).

Notice that the result above satisfies the strong-weak coupling duality
symmetry for the cross-noise. For example,
$S^{e}_{RI,RO}(\omega)=S^{q}_{RI,LO}(\omega)$ should be satisfied. Using the
result calculated above, $S^{e}_{RI,RO}(\omega)=\hbar |\omega|
(G_H-G^{e}_t)$, and $S^{q}_{RI,LO}(\omega)=\hbar |\omega| G^{q}_t$. Now,
$G_H-G^{e}_t=G^{e}_T=G^{q}_t$, so that, indeed,
$S^{e}_{RI,RO}(\omega)=S^{q}_{RI,LO}(\omega)$.

Finally, we can divide the cross-correlations by the equilibrium noise
$S^{(0)}(\omega)=\frac{\nu}{2\pi} |\omega|=\hbar |\omega|\; G_H$, and cast
the result in terms of normalized conductances $g_t=G_t/G_H$ and
$g_T=G_T/G_H$:

\begin{equation}
A_{aI,bO}=\delta_{a,b} -ab\ g_t =\cases
{
g_T ,&\mbox{$a=b$}\cr
g_t, &\mbox{$a\ne b$}\cr
}\ .
\end{equation}

\section{Auto-correlations}
\label{sec:current}
In this section, we will complete our calculation of the slope of the low 
frequency noise by finding the slope of the noise in the tunneling current 
$A_t$.  From $A_t$ and the cross-correlation calculated the previous section 
we can find all the other
correlators.  Since $A_t$ comes from the noise in the tunneling current, 
we will calculate directly the tunneling current-current correlation, given
by
\begin{equation}
S_t(\omega) = \int_{-\infty}^\infty dt \cos(\omega t) \;
\langle \{ I_t(t), I_t(0) \}\rangle \ ,
\end{equation}
where the tunneling operator is given by
\begin{equation}
I_t(t) = i \Gamma e^{i\omega_0 t} e^{i\sqrt{g}\phi(t,0)}
      - i \Gamma^* e^{-i\omega_0 t} e^{-i\sqrt{g}\phi(t,0)},
\label{eq:It}
\end{equation}
and $\phi(t,0) = \phi_R(t,0) + \phi_L(t,0)$. Notice that $S_t(\omega) =
S_t(-\omega)$, so that any term in $S_t$ that is linear in $\omega$ and
analytic must vanish.  Thus, at first order, the noise goes as $|\omega|$.
Also, because of the translational invariance of the correlator, we can
replace $\cos(\omega t)$ by $e^{i\omega t}$.

For comparison, we will once again need the tunneling current $I_t=\langle
I_t(t) \rangle$, where $I_t(t)$ is defined in Eq. (\ref{eq:It}). We can use
the expression for the tunneling current to obtain an expansion similar to
the one in Section \ref{sec:cross}.  In this case there is one ``physical" 
charge in 
the Coulomb gas which comes from the operator $I_t(t)$.  It is located at time
$t$ on the top branch and its charge is $q_0$, which can be $\pm 1$.  The
remaining inserted charges occur at times $t_i$ and can lie on either the top
or bottom branch, labeled by $\mu_i$, with $i = 1, \dots , 2n -1$.  They have
charges $q_i$ which are chosen so that the total charge (including $q_0$) is
zero.  With these definitions, the perturbation series for $I_t$ is
\begin{equation}
I_t = \sum_{n = 1}^\infty I^{(2n)},
\end{equation}
where
\begin{equation}
I^{(2n)} = {(-1)^n\ |\Gamma|^{2n}\over n!(n-1)!}
\sum_{\{\mu_i\}}\sum_{q_0 = \pm 1}
\int_{-\infty}^{\infty} \prod_{i=1}^{2n-1}\ \mu_idt_i 
\ \ \ e^{i\sum_{i=1}^{2n-1} \omega_0 q_i t_i}  e^{i \omega_0 q_0 t}
P_{+,{\mu_1},\dots,{\mu_{2n-1}}}(t, t_1,\dots,t_{2n-1}).
\end{equation}
In this equation, $P$ is defined as in Section \ref{sec:cross}.

For the noise, the Coulomb gas has a charge $q_0$ at $t\equiv t_0$ and a
charge $p_0$ at $s\equiv s_0$ which are both on the top branch.  Both of
these charges can be $\pm 1$.  The remaining charges $q_i$ are at positions
$t_i$ which can be on either branch, labeled by $\mu_i$, with $i = 1, \dots,
2n-2$.  The sum of all the charges must again be neutral. In particular, this
implies that the minimum number of inserted charges $n_{\rm min}(p_0,q_0)$ in
the perturbative expansion is $n_{\rm min}=0$ if $p_0$ and $q_0$ have
opposite signs, and $n_{\rm min}=2$ if $p_0$ and $q_0$ have the same sign.

The perturbation series for the noise in the tunneling current is then
\begin{eqnarray}
S_t(\omega) = 2\int_{-\infty}^{\infty} d(t-s) e^{i\omega(t-s)}
\sum_{\{\mu_i\}}
\sum_{q_0, p_0 = \pm 1} 
\sum_{ n = 1 + \frac{n_{\rm min}}{2}}^\infty
{(-1)^n\ |\Gamma|^{2n}\over n_+! n_-!}
\int_{-\infty}^{\infty} \prod_{i=1}^{2n-2}\ &\mu&_idt_i 
\ \ e^{i\sum_{i=1}^{2n-2} \omega_0 q_i t_i}  e^{i \omega_0 (q_0 t + p_0 s)}
\nonumber\\
\times &P&_{+,+,{\mu_1},\dots,{\mu_{2n-2}}}(t,s, t_1,\dots,t_{2n-2}),
\end{eqnarray}
where $n_+$ is the number of inserted charges that are positive and $n_-$ is
the number of negative inserted charges. The factor of $2$ in front of the
whole expression accounts for expanding the anti-commutator in the definition
of the noise into a time-ordered and an anti-time-ordered piece; for the
$|\omega|$ singularity, both contribute the same, hence we work with only the
time-ordered and include the factor of 2.

When $g > 1$ and $V$ is small, and also when $g <1$ and $V$ is large, the
leading contribution to the singularities should come from the configurations
where each of the charges at the $t_i$'s are close to either the charge at
$t$ or the charge at $s$.  For a particular configuration,
we will let $t_i$ for $i = 1, \dots, m$ be the
coordinates of the charges close to $t$, and $s_i$ for $i = 1, \dots, m'$ be
the coordinates of the charges close to $s$.  The charge at time $t_i$ has
charge $q_i$ and is on the branch labeled by $\mu_i$.  Similarly, the charge
at $s_i$ has charge $p_i$ and is on the branch $\nu_i$.  Then
\begin{eqnarray}
P_{+,+,\mu_1, \dots, \mu_{2n-2}}(t, s, t_1, \dots, t_{2n-2}) & = &
P_{+,\mu_1, \dots, \mu_m, +,\nu_1,\dots,\nu_{m'}}(t, t_1, \dots, t_{m}, s, s_1,
\dots, s_{m'}) \\
& = &
P_{+,\mu_1, \dots, \mu_m}(t, t_1, \dots, t_{m}) 
\ P_{+,\nu_1,\dots,\nu_{m'}}(s, s_1, \dots, s_{m'}) \nonumber\\
&\ &
\ \ \ \ \ \ 
\times \;R(t, t_1, \dots, t_m, s, s_1, \dots, s_m'),
\end{eqnarray}
where
\begin{equation}
R(t, t_1, \dots, t_m, s, s_1, \dots, s_m') = 
      e^{-g\sum_{i, j}q_ip_j
      [G_{{\mu_i}{\nu_j}}^{++}(t_i-s_j,0)
      +G_{{\mu_i}{\nu_j}}^{--}(t_i-s_j,0)]}.
\end{equation}
$R$ contains all the interactions between the charges in one multipole
and the charges in the other.  With the definition of the propagators $G$,
the expression for $R$ becomes
\begin{equation}
R(t, t_1, \dots, t_m, s, s_1, \dots, s_m') = 
       \prod_{i = 0}^m \prod_{j = 0}^{m'}
        \left(\delta + iK_{\mu_i\nu_j}(t_i - s_j)\right)^{2q_i p_j g}.
\end{equation}

If all the $t_i$ are shifted by $t$ and all the $s_i$ are shifted by $s$, then
the expression for $S(\omega)$ becomes
\begin{equation}
2\sum_{q_0, p_0 = \pm 1} \sum_{m, m' = 1}^\infty \int_{-\infty}^{\infty} 
   d(t-s) e^{i\omega (t-s)}  \tilde I_t^{(m)} \tilde I_t^{(m')} 
     R(t, t_1, \dots, t_m, s, s_1, \dots, s_m) e^{iQ\omega_0(t - s )},
\end{equation}
where $Q$ is the total charge in the multipole around $t$. $\tilde I_t^{(m)}$
and $\tilde I_t^{(m')}$ are defined similarly to $I_t^{(n)}$, but now the
charges do not have to sum to zero, and the integrals over $t_i$ and $s_i$ 
must also include the contribution from $R$.  It is straightforward 
to show that the
combinatorics in the equation for $I_t$ work out properly.  For example,
suppose there are $m_+$ positive charges in the multipole around $t$ and
$m_+'$ positive charges in the multipole around $s$.  In the $n$th term for
$S$, there are a total of $n_+$ positive inserted charges.  We must sum over
all the ways to pick the $m_+$ positive charges that are around $t$ from the
original $n_+$ positive charges, so the combinatorial factor becomes ${1
  \over n_+!}{n_+ \choose m_+} = {1 \over m_+! m_+'!}$.  Thus, the factorial
of the number of positive charges in $S$ is replaced by the product of the
factorials of positive charges in each $I_t$. The negative charges work
similarly.
 
Because $I_t^{(m)}$ and $I_t^{(m')}$ are independent of $t$ and $s$, and we
are left with evaluating the integral
\begin{equation}
R(\omega + Q\omega_0) = \int_{-\infty}^\infty d(t-s)
   R(t, t_1, \dots, t_m, s, s_1, \dots, s_m) e^{i(\omega +Q\omega_0)(t - s )}.
\end{equation}

First, we will assume the multipoles around $t$ and $s$ are both neutral.  In
that case, when $R$ is expanded out in powers of $1/(t -s)$, the first
two terms are given by
\begin{equation}
R(t, t_1, \dots, t_m, s, s_1, \dots, s_m') = 
     1 + {1 \over (t-s)^2} 2 g \sum_{i,j} q_i t_i p_j s_j.
\end{equation}
(We note that when the regulators are carefully kept track of, for each 
term in the sum the $1/(t-s)^2$ may be regulated differently.  However, since
we only want to find the leading singular behavior of the integral over
$t-s$, this short-distance behavior does not matter.)  Performing the
integral over $t-s$, we find that the singularity at $\omega = 0$ goes
as 
\begin{equation}
R_{\rm sing}(\omega) = -\pi |\omega| 2g \sum_{i,j}t_i q_i s_j p_j.
\end{equation}

If, instead, the multipoles around $t$ and $s$ are not neutral, but have
net charge $Q$, then $R(\omega + Q\omega_0) \propto 1/(t-s)^{2Q^2 g}$.
This will give a singularity in $S$ that goes as 
$|\omega \pm Q\omega_0|^{2Q^2 g -1}$.  Because this is smooth near 
$\omega = 0$ when $Q \ne 0$, this will not give any contribution to the
$\omega = 0$ singularity.  

Thus, the only contribution to the singularity at $\omega = 0$ comes from
$R_{\rm sing}(\omega)$.  From the equation for $I_t$, we note that
$\sum_{i=1}^{2n} t_i q_i I_t^{(n)}= {1\over i} {d \over d\omega_0} I_t^{n}$,
where the $t_i$ are understood to be inside the integral sign.
Using this equation and the expressions for $S_t(\omega)$ and $R_{\rm
  sing}(\omega)$, we find that the singular part of the noise near $\omega =
0$ is
\begin{equation}
S_t^{\rm sing} (\omega) = 
      4\pi g \left({d \over d\omega_0} I_t\right)^2 |\omega|.
\end{equation}
Now we can write the expresion above in terms of the differential conductance
$G_t=dI_t/dV$ and the Hall conductance $G_H=\nu e^2/h$, and set
$\omega_0=e^*V/\hbar$ to obtain
\begin{equation}
S_t^{\rm sing} (\omega) = 
     2 \;\hbar |\omega| \;\frac{G_t^2}{G_H}
\end{equation}
If we divide by $S^{(0)}(\omega)=\frac{\nu}{2\pi} |\omega|=\hbar |\omega|\;
G_H$, and write $g_t=G_t/G_H$, we find that the normalized slope of the
tunneling noise is
\begin{equation}
A_t = \frac{S_t^{\rm sing} (\omega)}{S^{(0)}(\omega)} = 2g_t^2,
\end{equation}
as we conjectured in Section \ref{sec-duality}.

\begin{multicols}{2}
\narrowtext

\section{Conclusion}
\label{sec:conclusions}

In this paper we present results on the low frequency part of the
noise spectrum for tunneling currents between fractional quantum Hall
edge states.  In addition to corrections to the classical shot noise
level $S=2e^* I$ due to generalized statistics of the quasiparticles,
the low frequency noise is not white or frequency independent. There
is cusp $\propto |\omega|$ in the spectrum, and we show in this paper
that the proportionality constant depends directly on the differential
conductance. Any anomalous scaling behavior on voltage $V$ or
tunneling amplitude $\Gamma$ are all contained in the implicit
dependence of the differential conductance on $V,\Gamma$.

In addition to studying the noise in tunneling and transmission
currents, we study the auto and cross-correlations of density-density
fluctuations between pairs of terminals in a four probe measurement
scheme. By looking into this larger class of correlations we can
identify different transformation laws under the duality symmetry
existent in the tunneling problem. More specifically, we can identify
(see Sections \ref{sec-results} and \ref{sec-duality}) correlations
that transform as either ``odd'' or ``anti-symmetric'', or ``even'' or
``symmetric'' quantities under duality.  The tunneling conductances is
an example of an ``odd'' quantity.

The transformation laws under duality are suggestive of relations
between the conductance and the slope of the noise correlations. We
pursue these relations via an ansatz, which we support through a
perturbative calculation.  We show that the perturbative expansions
for the slope of the noise spectrum at low frequencies matches term by
term the perturbative expansion series for differential conductances
(or squares of it). The equality makes use of the fact that the
tunneling Hamiltonian is a cosine potential at the point
contact. Therefore, our results are, in principle, particular to the
boundary sine-Gordon model. The equalities are independent of the
Luttinger parameter, so that one can perform the expansions around
either the electron or the quasiparticle tunneling limits, obtaining
the same relations to the differential conductances. The perturbative
expansions around these dual points are valid in complementary
regimes, so that the relations to the conductance hold for arbitrary
coupling.

Recently, the issue of finite frequency correlations has been studied
by Lesage and Saleur in Ref. \cite{LS3} using the thermal Bethe ansatz
and renormalized form factors. They also found that there is the
$\propto |\omega|$ singularity, and calculated the prefactor. However,
their $V$ and $\Gamma$ dependent prefactor does not coincide with the
square of the differential tunneling conductance. The source of
discrepancy remains unclear. Among the interesting and open questions
are those related to 1) dressing of the reflection coefficients in the
$S$-matrix 2) radius of convergence of the perturbative expansion. The
first issue arises because the dressing used for calculating the
conductance and the noise in Ref. \cite{LS3} are different.  The
rationale for the choice is that only states near the rapidity edge
(``Fermi'' level) of the solitons should participate in the low
$\omega$ noise. In the case of the conductance, all states under the
``Fermi'' sea contribute to the current. It is unclear if this
distinction is valid, since for the differential conductance only
states near the ``Fermi'' edge contribute. Of course, the system is
interacting, and the level positions depend self-consistently on the
occupation of other levels; but this rearrangement of levels should
also be expected when considering excitations on the scale $\omega$
for the noise correlations. The second issue concerns the validity of
the perturbative expansion to all orders. We believe the radius of
convergence of the series is finite, just as in the calculation of the
conductance \cite{FLS1,Weiss}. Moreover, the region of validity of one
expansion ends where the window for the dual expansion begins. Now,
although there is the duality symmetry in the boundary sine-Gordon
model, the theory in the infra-red fixed point requires
renormalization via neutral (density) counterterms \cite{CFW2}. At
first, one may expect that these counterterms would spoil the relation
between the slope of the noise spectrum and the conductance. However,
we have checked that at least the counterterm found in
Ref. \cite{CFW2} appear to leading order as $\omega^2$ corrections,
leaving the linear $\omega$ term unaffected.  The discrepancies
between our results and those of Ref. \cite{LS3} point to the need to
better understand the dynamical and non-equilibrium, $\omega,V \ne 0$,
aspects of the family of impurity problems described by the boundary
sine-Gordon model, as well the differences between the
Schwinger-Keldysh non-equilibrium formulation and the scattering
approach \cite{Zamolodchikov}.

Finally, we would like to put the contents of this paper in the
context of the recent advances in understanding the dual descriptions
of the boundary sine-Gordon problem \cite{Weiss,Fdual,FSdual}. An
integral representation was found for the conductance, in which the
dual expansions are controlled by the coupling dependent pole
structure and branch cut structure of the integrand. One should expect
that the slopes of the noise spectrum should have a similar integral
representation. Moreover, the different transformation laws (such as
the ``odd'' and ``even'' cases introduced here) should be manifest in
the form of the integral representations, and are natural quantities
for future studies.

\vspace{1cm}

\begin{center}
{\bf ACKNOWLEDGEMENTS} 
\end{center}

We would like to acknowledge many interesting discussions with Hubert
Saleur, who sent us the preprint of Ref. \cite{LS3} prior to
submission. Our work was motivated by these discussions, in an effort
to better understand the connection between the duality relations and
perturbative identities to the Bethe ansatz solution and the form
factors.  We would also like to thank Xiao-Gang Wen for many
enlighting discussions.  D.F. would like to acknowledge the Mary
Ingraham Bunting Institute which provided the opportunity to start 
this project.

\end{multicols}

\begin{thebibliography}{99}

\bibitem{Glattli}L.~Saminadayar, D.~C.~Glattli, Y.~Jin and B.~Etienne,
Phys.\ Rev.\ Lett.\ {\bf 79}, 2526 (1997).

\bibitem{Reznikov1}R.~de Picciotto, M.~Reznikov, M.~Heiblum, V.~Umansky,
G.~Bunin and D.~Mahalu, Nature {\bf 389}, 162 (1997).

\bibitem{K&F-noise}C.~L. Kane, and M.~P.~A.~ Fisher, 
Phys.\ Rev.\ Lett.\ {\bf 72}, 724 (1994).

\bibitem{FLS1}P.~Fendley, A.~W.~W.~Ludwig and H.~Saleur, 
Phys.\ Rev.\ Lett.\ {\bf 75}, 2196 (1995).

\bibitem{FS2}P.~Fendley and H.~Saleur, Phys.\ Rev.\ B {\bf 54}, 10845 (1996).

\bibitem{LS3}F.~Lesage and H.~Saleur, Nuc. Phys. B {\bf 493}, 613 (1997).

\bibitem{CFW1} C. de C. Chamon, D. E. Freed, and X. G. Wen, Phys.
Rev. B {\bf 51}, 2363 (1995).

\bibitem{CFW2}C.~de C.~Chamon, D.~E.~Freed, and X.~G.~Wen, Phys.\ Rev.\ B
{\bf 53} 4033, (1996) and references therein.

\bibitem{Weiss} U. Weiss, Solid State Comm. {\bf 100}, 281 (1996).

\bibitem{Martin} For a recent work on the zero-frequency shot noise
level for particles with generalized exclusion statistics, scattering
of an energy independent barrier, see S. B. Isakov, T. Martin, and
S. Ouvry, cond-mat/9811391.

\bibitem{Reznikov2} M. Reznikov, M. Heiblum, H. Shtrikman and D. Mahalu,
Phys. Rev. Lett. {\bf 75}, 3340 (1995).

\bibitem{Shiba} H. Shiba, Prog. Theor. Phys. {\bf 54}, 967 (1975).

\bibitem{Sassetti&Weiss} M. Sassetti and U. Weiss, Phys. Rev. Lett. {\bf
65}, 2262 (1990).

\bibitem{Zamolodchikov} V. Bazhanov, S. Lukyanov, and
A. Zamolodchikov, hep-th/9812247.

\bibitem{Fdual} P.~Fendley, hep-th/9804108.

\bibitem{FSdual} P.~Fendley and H.~Saleur, cond-mat/9804173.


\end{thebibliography}
\end{document}